\documentstyle[aps,bezier,epsfig,amsmath,amssymb]{revtex}
\begin{document}

\newcommand{\be}{\begin{eqnarray}}
\newcommand{\ee}{\end{eqnarray}}
\newcommand{\beq}{\begin{equation}}
\newcommand{\eeq}{\end{equation}}
\newcommand{\xx}{\begin{eqnarray*}}
\newcommand{\yy}{\end{eqnarray*}}
\newcommand{\nn}{\nonumber}
\newcommand{\Vol}{{\rm Vol}}
\newcommand{\sign}{{\rm sign}}
\newcommand{\tr}{{\rm Tr}}
\def\cut{{}\hfill\cr \hfill{}}

\title{Reply to Comment on `` Universal Fluctuations in Correlated
Systems''}

\author{ S.T. Bramwell$^1$,K. Christensen{$^2$}, J.-Y. Fortin$^{3\ast}$,
P.C.W. Holdsworth$^4$, H.J. Jensen$^5$, S. Lise{$^5$}, J.M.
L\'opez{$^{5\dagger}$}, M. Nicodemi{$^5$}, J.-F. Pinton$^4$, M.
Sellitto$^4$}

\address{$^1$ Department of Chemistry, University College London, 20
Gordon
Street, London, WC1H~0AJ, United Kingdom. \\
{$^2$}Blackett Laboratory, Imperial College,
Prince Consort Road, London SW7 2BZ United Kingdom.\\
$^3$ Department of Physics, University of Washington, Box
351560,Seattle, WA 98195-1560, USA. \\
$^4$ Laboratoire de Physique, Ecole Normale Sup\'erieure,
46 All\'ee d'Italie, F-69364 Lyon cedex 07, France.\\
$^5$ Department of Mathematics, Imperial College London, London,
SW7,United Kingdom.}

\maketitle


\medskip

\medskip

\noindent{\Large\bf Bramwell et al. reply}

\medskip

\medskip

Watkins Chapman and Rowlands claim to show that the deviation from
the FTG distribution, due to the correlations between random
variables introduced in the extremal statistics problem
in~\cite{1} is the result of the slow convergence of the PDF for
extreme values, with system size, for effective Gaussian
variables. We think this result is correct, but it must be put in
the context of  recent developments in the field.

Our original motivation for studying correlated extremal
statistics explicitly excludes the kind of slow relaxation towards
an asymptotic, or thermodynamic limit function discussed
in~\cite{2} and~\cite{3}: The PDF for order parameter fluctuations
in the low temperature phase of the 2D-XY model (BHP) is a
thermodynamic limit function that is different to the FTG
distribution. The model studied is diagonalisable into
statistically independent variables, for which the dispersion in
amplitudes diverges with system size. The PDF for the extreme
(largest) value of these variables is the FTG distribution not the
BHP distribution~\cite{4}. Conclusion: simple extremal statistics
do not explain the observed results. If extremal statistics are
relevant then they must apply to more complex (correlated) many
body objects, rather than the statistically independent variables
of the problem. If this is not the case then extremal statistics
are irrelevant for at least this model correlated system.

Our first attempt to look at the extremal statistics of correlated
random variables was the model presented in~\cite{1} and discussed
in the Comment. The authors are probably correct to conclude that
the main effect comes from finite size corrections, rather than
the correlations introduced in the model. This can therefore be
classified as "weak correlation" and is the extreme value
equivalent of introducing a finite correlation length in a
thermodynamic system and then taking the thermodynamic limit. In a
strong correlation limit one would expect deviations from FTG to
remain on taking the limit $N\rightarrow\infty$. This scenario has
been seen in detail in reference~\cite{5}, where the extreme value
for avalanche sizes in the Sneppen depinning model is seen to
follow the BHP distribution over a large range of time and length
scales. We also note that renormalization group analysis of
extreme value statistics for long range correlated signals shows
that the tail of the resulting distribution renormalizes from the
$\exp(-y)$ asymptote of the FTG distribution to
$y\exp(-y)$~\cite{6}, in agreement with the exact asymptotic form
for the BHP function~\cite{2}. The results of~\cite{5}
and~\cite{6} go somewhere towards confirming our hypothesis, first
proposed in~\cite{1}, that deviations to the FTG distribution
introduced by correlations could provide the desired link between
extremal statistics and the fluctuations of a global quantity in
such correlated systems. We suggest that the results presented
here (and therefore in \cite{1}) and in reference~\cite{3} pass
asymptotically close to this correlated regime.

Finally we remark that, even in the case where global fluctuations
{\it are} described by the FTG distribution, as in ref~\cite{7}
(see also ref.~\cite{8}), any connection with extremal statistics
remains unproven. The global quantity is a sum over a macroscopic
number of elements, while extremal statistics would require the
selection of the biggest element by a ``Maxwell's demon''. The
connection between these processes is an open problem. The authors
of this comment have presented arguments based on slow relaxation
towards the FTG distribution for Gaussian variables~\cite{3} which
could be a step in the right direction. This is a useful
contribution but the arguments presented here and in the
references below, show that it is far from the complete picture
and that it would be quite wrong to infer that deviation from the
FTG distribution in the ensemble of systems discussed is
generically due to finite size corrections to the thermodynamic
limit. More work needs to be done.

\end{document}